%% file: main.tex
\documentclass[]{spie}  

\usepackage{emmt-unicode}
\usepackage{emmt-codeblock}
\input{emmt-macros}
\usepackage{stmaryrd}
\usepackage[unicode=true, colorlinks=true, allcolors=blue]{hyperref}

\title{Beyond FRiM, ASAP: A family of Sparse Approximation for covariance matrices and Preconditioners}

\author[a]{Éric Thiébaut}
\author[a]{Michel Tallon}
\author[a]{Samuel Thé}
\author[b]{Loïc Denis}
\affil[a]{Univ. Lyon, Univ. Lyon 1, ENS de Lyon, CNRS, Centre de Recherche Astrophysique de Lyon UMR5574, F-69230, Saint-Genis-Laval, France}
\affil[b]{Univ. Lyon, UJM-Saint-Etienne, CNRS, Institut d’Optique Graduate School, Laboratoire Hubert Curien UMP 5516, F-42023, Saint-Étienne, France}

\authorinfo{Further author information: E-mail: eric.thiebaut@univ-lyon1.fr}

\newcommand*{\model}[1]{\widetilde{#1}}

\newcommand*{\mean}[1]{\overline{#1}}

\newcommand*{\Distance}{\mathcal{D}}

\newcommand*{\Potential}{\mathcal{V}}
\newcommand*{\Predicate}{\mathcal{P}}
\newcommand*{\bigO}{\mathcal{O}}
\newcommand*{\Id}{\M{I}}

\begin{document}
\maketitle

\begin{abstract}
The FRiM fractal operator belongs to a family of operators, called ASAP,
defined by an ordered selection of nearest neighbors. This generalization
provides means to improve upon the good properties of FRiM. We propose a fast
algorithm to build an ASAP operator mimicking the fractal structure of FRiM for
pupils of any size and geometry and to learn the sparse coefficients from
empirical data. We empirically show the good approximation by ASAP of
correlated statistics and the benefits of ASAP for solving phase restoration
problems.
\end{abstract}


\section{OUTLINE}
\label{sec:outline}

The FRiM\cite{Thiebaut_Tallon-2010-FRiM} fractal operator was proposed to
approximate the covariance matrix of the wavefront in phase reconstruction
algorithms and turned out to also be a fast and effective pre-conditioner for
these methods leading to a total computational burden scaling only as the size
of the reconstructed wavefront. This opens the possibility of real-time optimal
wavefront reconstruction by iteratively solving the inverse
problem\cite{Bechet_et_al-2007-Vancouver}. Yet FRiM intrinsic limitations can
be overstepped. Indeed, in Section~\ref{sec:generalizing-FRiM}, we demonstrate
that the FRiM fractal operator can be seen as a pivoted sparse Cholesky
factorization of the precision matrix of the wavefront. FRiM’s operator is then
a specific instance of a family of operators, called ASAP, which combine a
permutation and a sparse selection of nearest neighbors. In
Section~\ref{sec:building}, we propose an algorithm to build a sparse ordered
hierarchy for ASAP mimicking the fractal structure of FRiM for any size and
geometry (not just $(2^q+1)×(2^q+1)$ samples).  In Section~\ref{sec:learning},
we show that learning the coefficients of an ASAP approximation from empirical
data has a closed-form solution that is separable (and thus trivial to
parallelize). In Section~\ref{sec:results}, we present empirical results
showing the good approximation by ASAP of correlated statistics.

\section{GENERALIZING THE FRIM MODEL}
\label{sec:generalizing-FRiM}

A turbulent wavefront $\V x \in \Reals^n$ is a centered Gaussian variable.  A
general method to generate such a random vector consists in applying a linear
transform $\M K \from \Reals^n \to \Reals^n$ to a random vector $\V u \in
\Reals^n \sim \mathcal{N}\Paren{\V 0, \M I}$ following a standard normal
distribution ($\M I$ is the identity matrix of suitable size):
\begin{equation}
  \label{eq:random-node}
  \V x = \M K⋅\V u.
\end{equation}
The resulting random vector $\V x$ follows a centered Gaussian distribution of
covariance:
\begin{equation}
  \label{eq:covariance}
  \model{\M C} \bydef \Cov{\V x} = \M K⋅\Cov{\V u}⋅\M K\T = \M K⋅\M K\T,
\end{equation}
since $\Cov{\V u} = \Id$ and where exponent ${}\T$ denotes transposition.
Using the Cholesky decomposition of the true covariance $\M C = \M K⋅\M K\T$ or
the Karhunen-Loève transform \cite{Roddier-1990-wavefront_simulation} yields an
operator $\M K$ which implements the exact statistical model: $\model{\M C} =
\M C$. However, $\M K$ has up to $n\,(n+1)/2 = \bigO(n^2)$ non-zero entries with
the Cholesky decomposition, or $n^2$ if the Karhunen-Loève transform is used,
computing these entries takes $\bigO(n^3)$ numerical operations, and applying
$\M K$ takes $\bigO(n^2)$ operations.  This may be prohibitive, both in terms
of memory and of computational burden. In the case of random wavefronts caused
by the turbulence, Lane \emph{et al.}\cite{Lane_et_al-1992-Kolmogorov} proposed
a simulation method inspired by the fractal structure of turbulent wavefronts
and which is much faster than $\bigO(n^2)$. The fractal generator starts by a
$2\times2$ grid of random values and, at each stage, inserts new random values
so as to produce a refined grid whose step size is one half that of the
previous grid.  A given random value in the refined grid is generated by
combining an independent random variable and the values of, say, $m$ nearby
nodes in the coarser grid.  After $p$ iterations of the algorithm, the
generated random map has a size of $(2^{p}+1)\times(2^{p}+1)$ nodes.  As a
result, applying such an algorithm has a complexity of $\bigO(m\,n)$ to
generate $n$ correlated random variables.  When $m \ll n$, the fractal method
is very competitive even though the covariance of the generated variables is
not exactly the true one: $\model{\M C} \approx \M C$.  The approximation may
be improved by increasing the number $m$ of involved nearby
nodes\cite{Lane_et_al-1992-Kolmogorov}.  From Eq.~(23) of the FRiM
paper\cite{Thiebaut_Tallon-2010-FRiM}, the effect of the fractal operator $\M
K$ can be summarized by the following pseudo-code for computing $\V x = \M K⋅\V
u$:
\begin{equation}
  \codeBlock{
    \codeForL{i = 1, 2, \ldots, n}{
      x_{p_i} = A_{i,i} \, u_{p_i}
      + \sum\nolimits_{j = 1}^{i - 1} A_{i,j}\,x_{p_{j}}
    }
  }
  \label{eq:FRiM-direct-K}
\end{equation}
where $p_{i}$ is the index of the $i$-th node in the fractal hierarchy while
$A_{i,j}$ (for $j = 1,\ldots,i$) are precomputed coefficients.  The vector $\V
p =(p_1,p_2,\ldots,p_n)\T \in \IntRange{1,n}^n$ collecting these ordered
indices implements a permutation of indices $\IntRange{1,n}$ to re-order the
nodes from the largest to the smallest sampling scales while the doubly-indexed
pre-computed coefficients $A_{i,j}$ define a matrix $\M A \in \Reals^{n×n}$.
In the above pseudo-code, the nodes at scales smaller than the one considered
are not involved.  This is the same as if $A_{i,j} = 0$ when $j > i$, hence $\M
A$ is \emph{lower triangular}.  Furthermore, in FRiM algorithm, at most the $m$
previous sorted nodes $j$ that are the nearest to the $i$-th sorted node are
involved, hence $\M A$ is a lower triangular sparse matrix with at most
$\min(m+1,i)$ nonzero coefficients in its $i$-th row.

Denoting by $\M P$ the permutation matrix defined by the vector $\V p$ of
permuted indices and assuming that $\M R$ is a lower triangular matrix, then
the code to solve $\M P\T⋅\M R⋅\M P⋅\V x = \V u$ in $\V x$ is a \emph{pivoted
  backward back-substitution} which writes:
\begin{equation}
  \codeBlock{
    \codeForL{i = 1, 2, \ldots, n}{
      x_{p_i} = \Paren[\Big]{
        u_{p_i}
        - \sum\nolimits_{j = 1}^{i - 1} R_{i,j}\,x_{p_j}
      }/R_{i,i}
    }
  }
  \label{eq:pivoted-backward-backsubstitution}
\end{equation}
Noting that the inverse of a permutation matrix is its transpose and comparing
Eqs.~\eqref{eq:FRiM-direct-K} and \eqref{eq:pivoted-backward-backsubstitution}
directly shows that FRiM's fractal operator can be put in the following form:
\begin{equation}
  \M K = \Paren[\big]{\M P\T⋅\M R⋅\M P}^{-1} = \M P\T⋅\M R^{-1}⋅\M P,
\end{equation}
with $\M P$ the permutation matrix defined by the vector $\V p$ of permuted
indices and $\M R$ the lower triangular matrix whose entries are related to
those of $\M A$ by the following one-to-one relation:
\begin{equation}
  \label{eq:A<->R}
  R_{i,j} =
  \begin{cases}
    0 & \text{if $j > i$,}\\
    1/A_{i,i} & \text{if $j = i$,}\\
    -A_{i,j}/A_{i,i} & \text{if $j < i$,} \\
  \end{cases}
  \quad\Longleftrightarrow\quad
    A_{i,j} = \begin{cases}
    0 & \text{if $j > i$,}\\
    1/R_{i,i} & \text{if $j = i$,} \\
    -R_{i,j}/R_{i,i} & \text{if $j < i$.} \\
  \end{cases}
\end{equation}
for all $(i,j) \in \IntRange{1,n}^2$.

The covariance and precision matrices of the random vectors generated by FRiM
are:
\begin{align}
  \model{\M C} &\bydef \Cov{\V x} = \M K⋅\M K\T
  = \M P\T⋅\M R^{-1}⋅\M R\mT⋅\M P,
  \label{eq:C-model}\\
  \model{\M W} &\bydef \Cov{\V x}^{-1} = \M K\mT⋅\M K^{-1}
  = \M P\T⋅\M R\T⋅\M R⋅\M P.
  \label{eq:W-model}
\end{align}
These expressions show that FRiM's model in Eq.~\eqref{eq:W-model} of the
wavefront statistics amounts to approximating the true precision matrix $\M W =
\M C^{-1}$ of the wavefront by a \emph{pivoted sparse Cholesky
  decomposition}\cite{Bjorck-1996-least_squares,
  Higham-2009-Cholesky_factorization}.  Note that the inverse of a triangular
matrix being also triangular, FRiM's approximation in Eq.~\eqref{eq:C-model} of
the true covariance $\M C$ of the wavefront also takes the form of a pivoted
Cholesky decomposition although not necessarily sparse.

We propose to generalize the FRiM model by considering any approximation
$\model{\M W}$ of the precision matrix and corresponding approximation
$\model{\M C}$ of the covariance given by:
\begin{equation}
  \M C^{-1} \approx \model{\M W} = \M P\T⋅\M R\T⋅\M R⋅\M P
  \quad\Longleftrightarrow\quad
  \M C \approx \model{\M C} = \M P\T⋅\M R^{-1}⋅\M R\mT⋅\M P
  \label{eq:ASAP-family}
\end{equation}
with $\M P$ a permutation matrix and $\M R$ a sparse triangular matrix.
Equation~\eqref{eq:ASAP-family} defines \emph{A family of Sparse Approximation
  for covariance matrices and Preconditioners} which we call ASAP for short.

Finding the best ASAP approximation for, say, a given level of sparsity is a
very difficult problem.  In the following, we split this problem in two
sub-problems: (i) the construction of the sparse structure (including the
permutation) and (ii) the computation of the entries of the sparse matrix $\M
R$, the so-called \emph{structural non-zeros}.

\section{BUILDING THE SPARSE STRUCTURE}
\label{sec:building}

The sparse structure of an ASAP factor $\M P\T⋅\M R⋅\M P$ is defined by the
ordering of the nodes and, for each node, the list of connected nodes.  The
former is specified by the permutation $\M P$ or equivalently by the vector $\V
p$ of permuted indices.  The latter can be represented by:
\begin{equation}
  \Set{S}_i = \Brace[\big]{j \in \IntRange{1,n} \SuchThat (i,j) \in \Set{S}}
\end{equation}
the list of the nodes $j$ which are connected to the node $i$.  Here $\Set{S}$
denotes the sparsity pattern of the matrix $\M R$ which is the set of pairs
$(i,j)$ of row indices $i$ and column indices $j$ of the entries of $\M R$ not
forced to be zero, the so-called \emph{structural non-zeros} of $\M R$.
The level of sparsity is measured by $\mean{m}$ the mean number of connected
nodes per node.

In the following, we propose different possibilities to build the sparse
structure which are not limited to wavefronts of size $(2^q+1)×(2^q+1)$ as in
FRiM.

\subsection{Nearest neighbors without pivoting}
\label{sec:simple-nearest-neighbors}

A simple approach to define an ordered sparse structure is to assume no
permutations and use a \emph{predicate function} to determine whether a given
node is connected to another preceding node.  In this case, $\M P = \M I$ or
equivalently $\V p = (1,2,\ldots,n)\T$ and the structural non-zeros of node $i$
are given by:
\begin{equation}
  \Set{S}_i = \Brace[\big]{
    j\in\IntRange{1,n} \Given j ≤ i, \Predicate(i,j) = \mathtt{true}
  }
  \label{eq:simple-predicate-subset}
\end{equation}
where the condition $j ≤ i$ imposes that $j$ be a preceding node of $i$ or the
node itself (there are no permutations) and $\Predicate(i,j)$ is the predicate
function which yields $\mathtt{true}$ if $j$ is connected to $i$ and
$\mathtt{false}$ otherwise.  Note that since $i \in \Set{S}_i$ for each node
$i$, the condition $\Predicate(i,i) = \mathtt{true}$ must hold for each node
$i$.

\begin{figure}
  \centering
  \includegraphics[width=60mm]{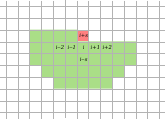}
  \caption{Example of nearest neighbors selection in 2-D with $\rho = 4$.  The
    green cells indicate the parent nodes $j \in \Set{S}'_i$ of the node $i$
    drawn in red, $s$ is the \emph{stride} that is the number of nodes per line
    of the grid.}
  \label{fig:nearest}
\end{figure}

In order to control the level of sparsity, we propose to choose to connect the
nearest preceding neighbors of a node and use the predicate:
\begin{equation}
  \Predicate(i,j) =
  \begin{cases}
    \mathtt{true} & \text{if $\Distance(i,j) ≤ \rho$} \\
    \mathtt{false} & \text{else} \\
  \end{cases}
  \label{eq:nearest-neighbors-predicate}
\end{equation}
where $\Distance(i,j)$ yields the distance between nodes $i$ and $j$ and $\rho
≥ 0$ is a distance selection threshold.  The smaller $\rho$ the sparser the
factor $\M R$.  In the limit $\rho \rightarrow 0$, $\Set{S}_i \rightarrow
\Brace{i}$ which amounts to having $\M R$ diagonal.  Figure~\ref{fig:nearest}
shows a node $i$ and its parents $\Set{S}'_i = \Set{S}_i \backslash \Brace{i}$
assuming a Euclidean distance on a 2-D grid of nodes and a threshold $\rho = 4$
in units of the grid step size.

\subsection{Nearest neighbors with given ordering}
\label{sec:ordered-nearest-neighbors}

We next consider an ordered sparse structure based on a given permutation $\M
P$ and connecting the nearest preceding nodes to each node.  The maximal number
$m$ of structural non-zeros per node is used to control the sparsity level.  In
order to build such a sparse structure, we use an auxiliary matrix $\M J \in
\Integers^{m×n}$.  At any stage of the construction, the column $i$ of $\M J$
stores the list of the nodes connected to $i$.  As there are at most $m$ such
nodes, $\M J$ is of size $m×n$.  To indicate unset entries in the auxiliary
index matrix $\M J$, we use any invalid node number, say $j = -1$.  As we want
to select the nearest preceding nodes, we use another auxiliary matrix $\M D
\in \Reals^{m×n}$ to keep track of the distances between connected nodes.  In
other words, the entries of the auxiliary matrices are such that $J_{\ell,i} =
j$ with $j$ the $\ell$-th node connected to node $i$ and $D_{\ell,i} =
\Distance(i,j)$ the distance between these two nodes.  We require that the
following properties hold for the \emph{distance} $\Distance$:
\begin{subequations}
  \label{eq:dist-props}
  \begin{align}
    \label{eq:D(i,j)≥0}
    &\forall (i,j),\ \Distance(i,j) ≥ 0,\\
    \label{eq:D(i,i)=0}
    &i = j \quad\Longleftrightarrow\quad \Distance(i,j) = 0,\\
    \label{eq:D(i,j)>0}
    &i \not= j \quad\Longleftrightarrow\quad \Distance(i,j) > 0,\\
    \label{eq:D(i,-1)=inf}
    &\forall i,\ \Distance(i,-1) = +\infty.
  \end{align}
\end{subequations}
To simplify the code of the construction algorithm, we set the distances of
empty entries to $+\infty$ in the auxiliary distance matrix $\M D$ as stated by
Eq.~\eqref{eq:D(i,-1)=inf} and sort the entries in the columns of $\M J$ and
$\M D$ in order of non-increasing distances.  Putting all together, we require
that the following properties hold for the auxiliary matrices $\M J$ and $\M D$
at any stage of the construction algorithm:
\begin{subequations}
\begin{align}
  \label{eq:last-entry}
  &(J_{m,i},D_{m,i}) = (i,0), \\
  \label{eq:J_li}
  &J_{\ell,i} = \begin{cases}
    -1 & \text{if not yet set}\\
    j \in \IntRange{1,n}  & \text{otherwise and with $r_j ≤ r_i$}\\
  \end{cases}\\
  \label{eq:D_li}
  &D_{\ell,i} = \begin{cases}
    +\infty & \text{if $J_{\ell,i}$ not yet set}\\
    \Distance(i, J_{\ell,i}) & \text{otherwise}\\
  \end{cases}\\
  \label{eq:dist-order}
  &D_{1,i} ≥ D_{2,i} ≥ \ldots ≥ D_{m-1,i} >  D_{m,i} = 0\\
  \label{eq:min-dist}
  &\min_\ell D_{\ell,i} = D_{m,i} = 0\\
  \label{eq:max-dist}
  &\max_\ell D_{\ell,i} = D_{1,i} &\text{(note that $D_{1,i} > 0$ if $m > 1$)}
\end{align}
\end{subequations}
Since $i \in \Set{S}_i$ and $\Distance(i,i) = 0$ is the least possible
distance, the last entry in each column $i$ of the auxiliary matrices $\M J$
and $\M D$ is for $j=i$, this is reflected by Eq.~\eqref{eq:last-entry}.
Equation~\eqref{eq:dist-order} reflects the non-increasing distance ordering of
entries in the columns of the auxiliary matrices $\M J$ and $\M D$ and
Eqs.~\eqref{eq:min-dist} and \eqref{eq:max-dist} follows from this ordering and
from the other properties.  Equation~\eqref{eq:J_li} introduces the \emph{rank}
vector $\V r \in \IntRange{1,n}^n$ such that $r_i$ yields the rank of the node
$i$ in the chosen permutation.  Hence the constraints $r_j ≤ r_i$ in
Eq.~\eqref{eq:J_li} simply states that $j$ is a preceding node of $i$ or $i$
itself.  The following identities hold for rank and permutation vectors $\V r$
and $\V p$:
\begin{subequations}
\begin{align}
  \label{eq:rank-perm}
  &\forall k \in \IntRange{1,n},\ r_{p_k} = k,\\
  \label{eq:perm-rank}
  &\forall i \in \IntRange{1,n},\ p_{r_i} = i.
\end{align}
\end{subequations}
Remark that Eq.~\eqref{eq:rank-perm} can be used to build the rank vector $\V
r$ from the permutation vector $\V p$ and, conversely, Eq.~\eqref{eq:perm-rank}
can be used to build the permutation vector $\V p$ from the rank vector $\V r$.

We now have all necessary equations to describe
Algorithm~\ref{alg:build-struct-given-order} which implements the construction
of a sparse structure with a given ordering of the nodes and which retains at
most $m$ nearest preceding neighbors for each node.  After proper
initialization and at each stage of the construction,
Algorithm~\ref{alg:build-struct-given-order} picks the next node $j$ according
to the permutation order and, for each ensuing node $i$, checks whether $j$ is
closer to $i$ than any of its current nearest preceding neighbors; if this is
the case, the algorithm replaces the most remote node connected to $i$ by the
node $j$ while maintaining the ordering stated by Eq.~\eqref{eq:dist-order}.
This operation is performed by the $\mathtt{update}$ method implemented by
Algorithm~\ref{alg:update} and which mostly amounts to performing a \emph{sort
  by insertion} to maintain the distance ordering of the connected nodes.  Note
that this updating leave unchanged all connections recorded by the node $j$ and
its predecessors: at stage $k$, the structure is \emph{frozen} for the $k$
first nodes.  At the end of Algorithm~\ref{alg:build-struct-given-order}, the
list of structural non-zeros for each node $i$ is simply given by the valid
entries of the column $i$ of the auxiliary index matrix:
\begin{equation}
  \label{eq:extract-Si}
  \Set{S}_i = \Brace[\big]{
    J_{\ell,i} \SuchThat \ell \in \IntRange{1,m} \text{ and } J_{\ell,i} \not= -1
  }.
\end{equation}
It is important to realize that all the required properties must hold for the
algorithm to work correctly.  It may be checked that these properties are
maintained at all stages by the proposed method.

\begin{algorithm}
  \caption{Build structure for given permutation and distance (reference
    version).}

  \KwIn{Un-initialized auxiliary matrices $\M J \in \Integers^{m×n}$ and $\M D
    \in \Reals^{m×n}$, permuted indices $\V p
    \in \IntRange{1,n}^n$, and distance
    $\Distance\from\IntRange{1,n}\times\IntRange{1,n}\to \Reals_+ \cup
    \Brace{+\infty}$.}

  \KwOut{Initialized auxiliary matrices.}

  \BlankLine
  \CommentLine{Initialization.}
  \For{$i = 1,2,\ldots,n$}{
    \For{$\ell = 1,2,\ldots,m-1$}{
      $(J_{\ell,i},D_{\ell,i}) \gets (-1, +\infty)$
      \Comment*{see Eqs.~\eqref{eq:J_li} and \eqref{eq:D_li}}
    }
    $(J_{m,i},D_{m,i}) \gets (i,0)$ \Comment*{see Eq.~\eqref{eq:last-entry}}
  }
  \BlankLine
  \CommentLine{Construction.}
  \For{$k = 1,2,\ldots,n$}{
    $j = p_k$\\
    \For{$k' = k+1,k+2,\ldots,n$}{
      $i = p_{k'}$\\
      $d = \Distance(i,j)$\\
      \If{$d < D_{1,i} = \max_\ell D_{\ell,i}$}{
        $\mathtt{update}(\M J,\M D,i,j,d)$\\
      }
    }
  }
  \label{alg:build-struct-given-order}
\end{algorithm}

\begin{algorithm}
  \caption{$\mathtt{update}(\M J,\M D,i,j,d)$ procedure.}

  \KwIn{Auxiliary matrices $\M J \in \Integers^{m×n}$ and $\M D \in
    \Reals^{m×n}$, nodes $(i,j) \in \IntRange{1,n}\times\IntRange{1,n}$ such
    that $r_j < r_i$, and distance $d = \Distance(i,j)$ such that $d <
    D_{1,i} = \max_\ell D_{\ell,i}$.}

  \BlankLine

  $\ell \gets 1$\\
  \While{$\ell < m$ \KwAnd $d < D_{\ell+1,i}$}{
    $(J_{\ell,i},D_{\ell,i}) \gets (J_{\ell+1,i},D_{\ell+1,i})$\\
    $\ell \gets \ell + 1$\\
  }
  $(J_{\ell,i},D_{\ell,i}) \gets (j,d)$\\
  \label{alg:update}
\end{algorithm}

\subsection{Nearest neighbors with automatic ordering}
\label{sec:nearest-neighbors-auto-order}

In order to closely mimic the fractal structure of the ordering of nodes in
FRiM algorithm, we propose another way to build a sparse structure which
retains at most $m$ nearest preceding neighbors for each node but with an
ordering of nodes that is automatically built so that, at any stage of the
construction, the \emph{density} of the ordered nodes is as uniform as
possible.  Said otherwise, we want that the voids due to the unordered nodes to
all have approximately the same volume.

Using the rules and auxiliary matrices introduced in
Section~\ref{sec:ordered-nearest-neighbors} implementing such a construction is
pretty straightforward provided we have a mean to estimate the volume of the
voids due to unordered nodes.  To that end, we propose to relate the volume of
the void around a given unordered node $i$ to the following potential:
\begin{equation}
  v_i = \sum_{j \in \Set{S}'_i} \Potential(\Distance(i,j))
\end{equation}
where $\Set{S}'_i \bydef \Set{S}_i\backslash\Brace{i}$ is the list of neighbors
of the node $i$ (without $i$ itself) and $\Potential\from\Reals_+\to\Reals_+$
is a given potential function which is a decreasing function of the distance
between nodes.  Equipped with this potential, we can now state a simple rule to
select the next ordered node among the unordered ones: pick the one that sit in
the largest void according to its potential.  As the potential function
$\Potential$ is a decreasing function of the distance, then the unordered node
with the lowest potential is sitting in the largest void.  In practice and by
analogy with particle physics, we take the potential function to be the
reciprocal of the distance:
\begin{equation}
  \Potential(d) = 1/d,
\end{equation}
and thus select as the next ordered node the unordered node with the least
total potential.

\begin{figure}
  \centering
  \includegraphics[width=\textwidth]{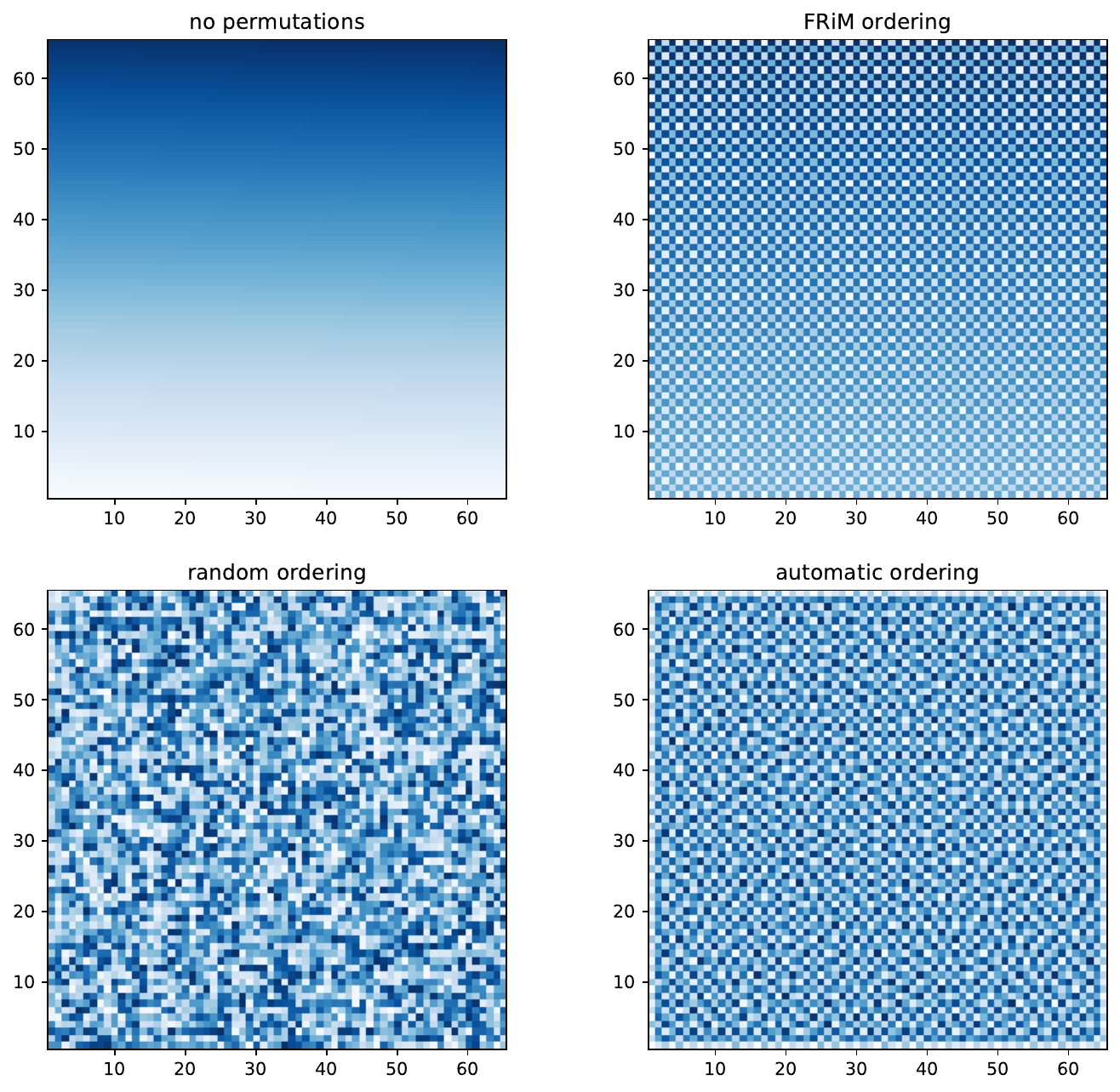}
  \caption{Ordering of $65×65$ nodes by different methods.  The first nodes in
    the ordering have the brightest shades of blue, the last nodes in the
    ordering have the darkest shades of blue.}
  \label{fig:orderings}
\end{figure}

Algorithm~\ref{alg:build-struct-auto-order} implements the proposed method
re-using most of Algorithm~\ref{alg:build-struct-given-order} and the
$\mathtt{update}$ method in Algorithm~\ref{alg:update}.  In addition to
auxiliary matrices $\M J \in \Integers^{m×n}$ and $\M D \in \Reals^{m×n}$,
distance and potential functions $\Distance$ and $\Potential$, and
un-initialized permutation vector $\V p \in \Integers^n$, the algorithm
requires to chose the first node to start with and maintain a set $\Set U$ of
unordered nodes.  On return of Algorithm~\ref{alg:build-struct-auto-order}, the
permutation vector $\V p$ is initialized and Eq.~\eqref{eq:extract-Si} can be
used to extract the structural non-zeros $\Set S_i$ for each node $i$.

\begin{algorithm}
  \caption{Build structure with automatic ordering (reference version).}

  \KwIn{Un-initialized auxiliary matrices $\M J \in \Integers^{m×n}$ and $\M D
    \in \Reals^{m×n}$, un-initialized permutation vector $\V p \in \Integers^n$,
    distance $\Distance\from\IntRange{1,n}\times\IntRange{1,n}\to \Reals_+ \cup
    \Brace{+\infty}$, potential function $\Potential\from\Reals_+\to\Reals_+$,
    initial node $j_\Tag{init}$, and workspace $\V v \in \Reals^n$.}

  \KwOut{Initialized auxiliary matrices $\M J$ and $\M D$, and initialized
    permutation vector $\V p$.}

  \BlankLine
  \CommentLine{Initialization.}
  \For{$i = 1,2,\ldots,n$}{
    \For{$\ell = 1,2,\ldots,m-1$}{
      $(J_{\ell,i},D_{\ell,i}) \gets (-1, +\infty)$
      \Comment*{see Eqs.~\eqref{eq:J_li} and \eqref{eq:D_li}}
    }
    $(J_{m,i},D_{m,i}) \gets (i,0)$ \Comment*{see Eq.~\eqref{eq:last-entry}}
  }
  $\Set{U} \gets \IntRange{1,n}$\\
  \BlankLine
  \CommentLine{Construction.}
  \For{$k = 1,2,\ldots,n$}{
      \uIf{$k = 1$}{
        $j \gets j_\Tag{init}$\\
      }
      \Else{
        $j \gets \argmin_{i\in\Set{U}}v_i$\\
      }
      $p_k \gets j$\\
      $\Set{U} \gets \Set{U} \backslash \Brace{j}$\\
      \For{$i \in \Set{U}$}{
        $d = \Distance(i,j)$\\
        \If{$d < D_{1,i} = \max_\ell D_{\ell,i}$}{
          $\mathtt{update}(\M J,\M D,i,j,d)$\\
          $v_i \gets \sum_{\ell=1}^{m-1} \Potential(D_{\ell,i})$\\
      }
    }
  }
  \label{alg:build-struct-auto-order}
\end{algorithm}

Figure~\ref{fig:orderings} shows the ranks of the nodes of a $65×65$ wavefront
for the different orderings studied in this paper.  With no permutations, the
ordering follows the lexicographic order and connected nodes are always at
similar short distances.  For the FRiM ordering nodes are ordered so that they
define a regular sampling at decreasing scales.  The resulting regular pattern
can be seen clearly.  The random ordering achieves a similar multi-scale
sampling but obviously with no regular pattern but rather with a tendency to
form clusters of voids.  The automatic ordering achieves a multi-scale sampling
but without the very regular pattern of FRiM and without the artifacts
(clusters and voids) of the random ordering.  From this simple analysis, we
expect the latter ordering method to produce the best results.

Algorithms~\ref{alg:build-struct-given-order} and
\ref{alg:build-struct-auto-order} implement \emph{reference versions} of the
construction methods which scale as $\bigO(n^2)$ or worse in terms of numerical
complexity.  These implementations may be improved to be much faster, describing
such improvements is beyond the scope of the present paper.

\section{LEARNING THE MODEL}
\label{sec:learning}

In the previous section we proposed different strategies to build an ordered
sparse structure which yields the permutation $\M P$ and the set $\Set S$ of
structural non-zeros of the factor $\M R$ (or $\M A$).  In order to compute the
values of the structural non-zeros, we propose to generalize FRiM's approach
for estimating the coefficients of $\M A$ in Eq.~\eqref{eq:FRiM-direct-K} to
any other sparse structure.

\subsection{FRiM's approach for computing the coefficients of $\M A$}
\label{eq:learn-A}

To simplify the developments, we introduce the permuted variables $\mathring{\V
  x} \bydef \M P⋅\V x$ and $\mathring{\V u} \bydef \M P⋅\V u$ (hence
$\mathring{x}_{i} = x_{p_i}$ for any index $i$).  Then, from the pseudo-code in
Eq.~\eqref{eq:FRiM-direct-K} and since variables are centered, the covariance
between the sample $\mathring{x}_i$ and the preceding sample $\mathring{x}_j$
is given by:
\begin{align}
  \Cov{\mathring{x}_i,\mathring{x}_j}
  &= \Avg[\Big]{
    \Paren[\Big]{
      A_{i,i}\,\mathring{u}_i
      + \sum\nolimits_{j' \in \Set{S}'_i} A_{i,j'}\,\mathring{x}_{j'}
    }\,\mathring{x}_{j}
  }
  = A_{i,i}\,\Avg[\big]{\mathring{u}_i\,\mathring{x}_{j}}
  + \sum\nolimits_{j'  \in \Set{S}'_i} A_{i,j'}\,
  \Avg[\big]{\mathring{x}_{j'}\,\mathring{x}_{j}}\notag\\
  &= \sum\nolimits_{j'  \in \Set{S}'_i} A_{i,j'}\,\mathring{C}_{j',j}
  = \V A_{i,\Set{S}'_i}⋅\mathring{\V C}_{\Set{S}'_i,j}
  \label{eq:cov(xi,xj)}
\end{align}
where $\Avg{\ldots}$ denotes expectation and where
$\Avg[\big]{\mathring{u}_i\,\mathring{x}_{j}} =
\Avg[\big]{\mathring{u}_i}\,\Avg[\big]{\mathring{x}_{j}} = 0$ because the
variables $\mathring{u}_i$ and $\mathring{x}_{j}$ are independent and centered.
Above, $\V A_{i,\Set{S}'_i}$ denotes the row vector which is the restriction to
$\Set{S}'_i = \Set{S}_i \backslash \Brace{i}$ of the $i$-th row vector of $\M
A$, that is the row vector of the off-diagonal structural non-zeros of the
$i$-th row of $\M A$.  The same conventions of notation apply for the column
vector $\mathring{\V C}_{\Set{S}'_i,j}$ built from $\mathring{\M C}$ the target
covariance matrix for the permuted variables $\mathring{\V x}$.  Requiring that
$\Cov{\mathring{x}_i,\mathring{x}_j} = \mathring{C}_{i,j}$ for all
$j\in\Set{S}'_i$ yields a system of linear equations whose solution are the
off-diagonal structural non-zeros of the $i$-th row of $\M A$:
\begin{align}
  \label{eq:FRiM-Aij}
  \mathring{\V C}_{i,\Set{S}'_i} =
  \V A_{i,\Set{S}'_i}⋅\mathring{\M C}_{\Set{S}'_i,\Set{S}'_i}.
\end{align}

FRiM also requires that the variance of $\mathring{x}_i$ be equal to the true
variance $\mathring{C}_{i,i}$ which amounts to having:
\begin{align*}
  \Var{\mathring{x}_i}
  &= \Avg*{
    \Paren[\Big]{
      A_{i,i}\,\mathring{u}_i
      + \sum\nolimits_{j  \in \Set{S}'_i} A_{i,j}\,\mathring{x}_{j}
    }^2
   }\notag\\
  &= A_{i,i}^2\,\Avg[\big]{\mathring{u}_i^2}
  + 2\,A_{i,i}\,\sum\nolimits_{j  \in \Set{S}'_i} A_{i,j}\,
  \Avg[\big]{\mathring{u}_i\,\mathring{x}_{j}}
  + \sum\nolimits_{j  \in \Set{S}'_i} \sum\nolimits_{j'  \in \Set{S}'_i}
  A_{i,j}\,A_{i,j'}\,\Avg[\big]{\mathring{x}_j\,\mathring{x}_{j'}}\notag\\
  &\approx A_{i,i}^2
  + \sum\nolimits_{j  \in \Set{S}'_i} \sum\nolimits_{j'\in\Set{S}'_i}
  A_{i,j}\,A_{i,j'}\,\mathring{C}_{j,j'},
\end{align*}
where the last right hand side come from
$\Avg[\big]{\mathring{u}_i\,\mathring{x}_{j}} = 0$ for any $j < i$ (as before),
from $\Avg[\big]{\mathring{u}_i^2} = \Var[\big]{\mathring{u}_i} = 1$, and from
the assumption that:
\begin{equation}
  \Avg[\big]{\mathring{x}_j\,\mathring{x}_{j'}} \approx \mathring{C}_{j,j'}.
\end{equation}
Note that this approximation would be exact if we were not imposing some level
of sparsity.  Now, using the fact that the coefficients $A_{i,j}$ for $j < i$
are such that the right hand side of Eq.~\eqref{eq:cov(xi,xj)} be equal to
$\mathring{C}_{i,j}$ yields:
\begin{align}
  \label{eq:FRiM-Aii}
  \Var{\mathring{x}_i}
  &= A_{i,i}^2 + \sum\nolimits_{j  \in \Set{S}'_i} A_{i,j}\,\mathring{C}_{j,i}
  = A_{i,i}^2 + \V A_{i,\Set{S}'_i}⋅\mathring{\V C}_{\Set{S}'_i,i}.
\end{align}
Beware that Eq.~\eqref{eq:cov(xi,xj)} cannot be used to further simplify the
last term of the above right hand side because Eq.~\eqref{eq:cov(xi,xj)} is for
$\mathring{C}_{j,j'}$ with both $j < i$ and $j' < i$.

Jointly solving Eqs~\eqref{eq:FRiM-Aij} and \eqref{eq:FRiM-Aii} generalizes the
computation of the coefficients of the fractal operator in original FRiM method
to any sparsity pattern.  It is however possible to obtain a closed-form
expression by solving these equations for $\M R$ instead of $\M A$ as considered
next.

\subsection{Learning of the coefficients of $\M R$ the FRiM way}
\label{eq:learn-R}

Combining Eq.~\eqref{eq:A<->R} with Eqs~\eqref{eq:FRiM-Aij} and
\eqref{eq:FRiM-Aii}, the structural non-zeros of $\M R$ are given by solving
the following set of equations:
\begin{align}
  \left\{\begin{array}{l}
  \V R_{i,\Set{S}'_i}⋅\mathring{\M C}_{\Set{S}'_i,\Set{S}'_i}
  + R_{i,i}\,\mathring{\V C}_{i,\Set{S}'_i} = \V 0\\
  \displaystyle
  \V R_{i,\Set{S}'_i}⋅\mathring{\V C}_{i,\Set{S}'_i}\T
  + R_{i,i}\,\mathring{C}_{i,i} - \frac{1}{R_{i,i}} = 0
  \end{array}\right.
  \quad\Longleftrightarrow\quad
  \V R_{i,\Set{S}_i}⋅\mathring{\M C}_{\Set{S}_i,\Set{S}_i}
  = \frac{1}{R_{i,i}}\,\V I_{i,\Set{S}_i}.
  \label{eq:solving-R-1}
\end{align}
This latter compact form expression combines all the structural non-zeros of
the row $i$ of $\M R$.  $\V I_{i,\Set{S}_i}$ is the restriction of the row $i$
of the identity $\M I$ matrix to the column indices $\Set{S}_i$ of the
structural non-zeros of the row $i$ of $\M R$.  In other words, $\V
I_{i,\Set{S}_i}$ is a row vector which is zero everywhere except in the entry
corresponding to the diagonal where it is one.  From Eq.~\eqref{eq:solving-R-1}
follows immediately that:
\begin{equation}
  \label{eq:solving-R-2}
  \V R_{i,\Set{S}_i} = \frac{1}{R_{i,i}}\,\V I_{i,\Set{S}_i}⋅
  \Paren[\big]{\mathring{\M C}_{\Set{S}_i,\Set{S}_i}}^{-1},
\end{equation}
because $\mathring{\M C}_{\Set{S}_i,\Set{S}_i}$, being the restriction to
  $\Set{S}_i$ of the positive definite matrix $\mathring{\M
    C}_{\Set{S}_i,\Set{S}_i}$, it is invertible.  To estimate the diagonal
  term, we note that $R_{i,i} = \V R_{i,\Set{S}_i}⋅\V I_{i,\Set{S}_i}\T$ and
  combining this with Eq.~\eqref{eq:solving-R-2}, we obtain:
\begin{displaymath}
  R_{i,i}
  = \V R_{i,\Set{S}_i}⋅\V I_{i,\Set{S}_i}\T
  = \frac{1}{R_{i,i}}\,\V I_{i,\Set{S}_i}⋅
  \Paren[\big]{\mathring{\M C}_{\Set{S}_i,\Set{S}_i}}^{-1}⋅\V I_{i,\Set{S}_i}\T.
\end{displaymath}
Since $\mathring{\M C}_{\Set{S}_i,\Set{S}_i}$ is positive definite, its inverse
is also positive definite and thus $\V I_{i,\Set{S}_i}⋅\mathring{\M
  C}_{\Set{S}_i,\Set{S}_i}^{-1}⋅\V I_{i,\Set{S}_i}\T > 0$ and:
\begin{displaymath}
  \label{eq:diag-term-R}
  R_{i,i} = \pm\sqrt{
    \V I_{i,\Set{S}_i}⋅
    \Paren[\big]{\mathring{\M C}_{\Set{S}_i,\Set{S}_i}}^{-1}⋅
    \V I_{i,\Set{S}_i}\T
  },
\end{displaymath}
are the 2 possible solutions for the diagonal term.  Finally, the structural
non-zeros of the row $i$ of $\M R$ have the following closed-form expression:
\begin{equation}
  \label{eq:closed-form-R}
  \boxed{
    \V R_{i,\Set{S}_i} = \frac{
      \pm\V I_{i,\Set{S}_i}⋅\Paren[\big]{\mathring{\M C}_{\Set{S}_i,\Set{S}_i}}^{-1}
    }{
      \sqrt{
        \V I_{i,\Set{S}_i}⋅
        \Paren[\big]{\mathring{\M C}_{\Set{S}_i,\Set{S}_i}}^{-1}⋅
        \V I_{i,\Set{S}_i}\T
      }
    }
  }
\end{equation}
Whatever the choice of the sign, $\M R\T⋅\M R$ yields the same approximation of
$\mathring{\M C}^{-1}$.

Looking at Eq.~\eqref{eq:closed-form-R} for $\M R$, it is clear that, unlike
the Cholesky decomposition, the computation of the coefficients of $\M R$ can
be performed independently for each row of $\M A$.  The same remark applies for
$\M A$ considering Eqs~\eqref{eq:FRiM-Aij} and \eqref{eq:FRiM-Aii}.
Hence, these computations can be easily accelerated on a multi-core or
multi-processor computer by a trivial parallelization\footnote{problems with
  such structure are called \emph{embarrassingly parallel} to emphasize that
  they are suitable for massive parallelization}.

\begin{figure}
  \centering
  \includegraphics[width=0.47\textwidth]{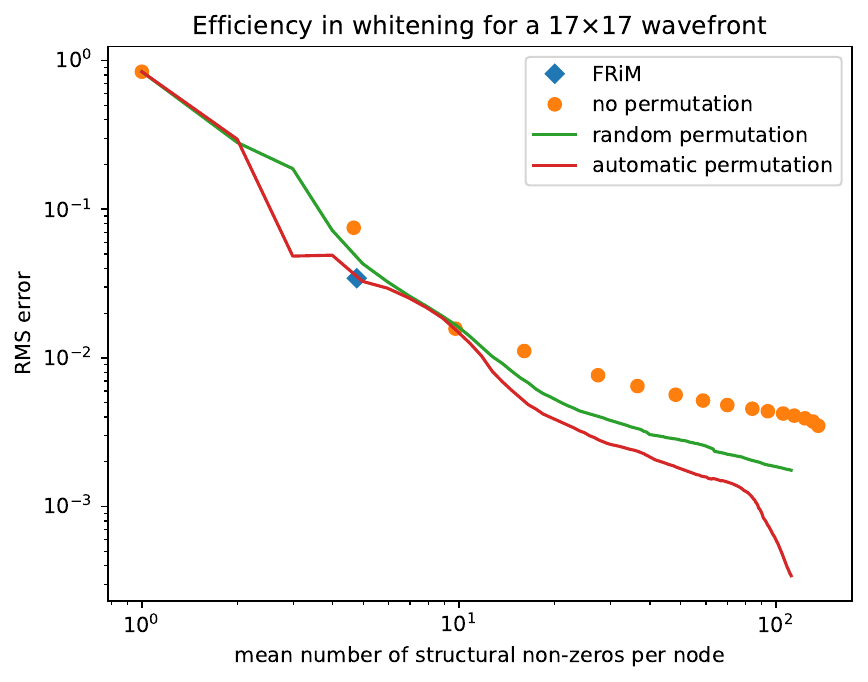}\hfill%
  \includegraphics[width=0.47\textwidth]{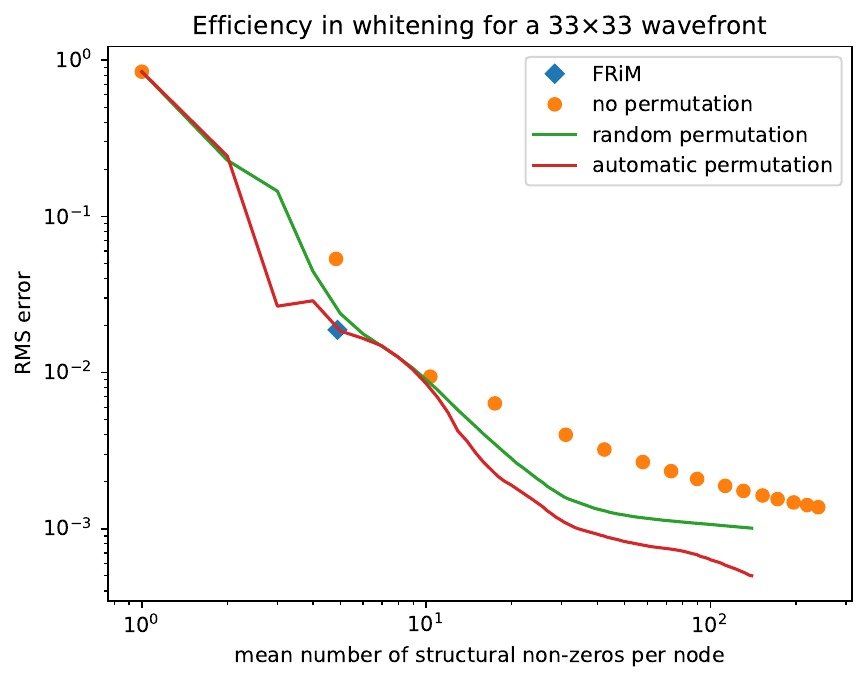}
  \includegraphics[width=0.47\textwidth]{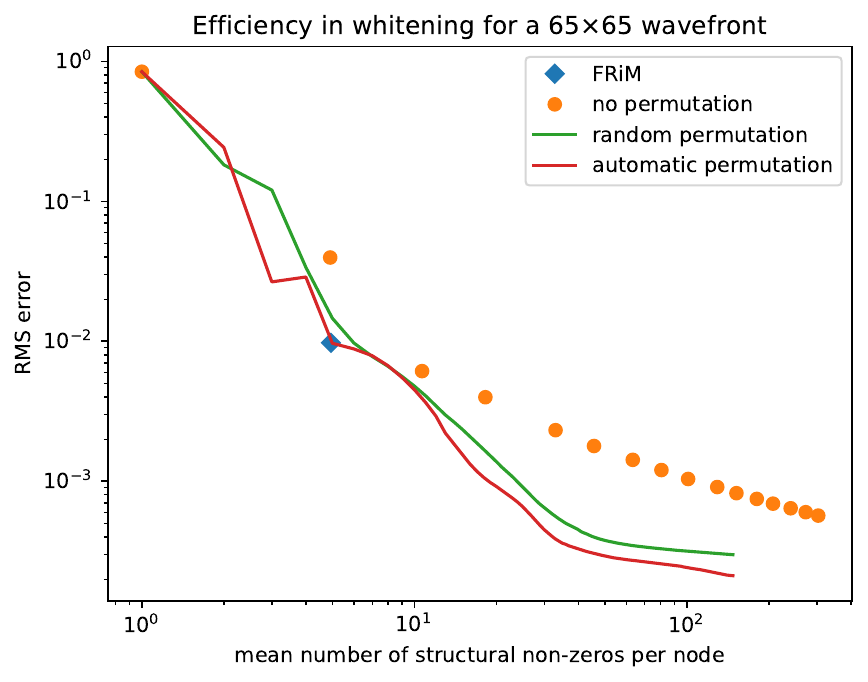}\hfill%
  \includegraphics[width=0.47\textwidth]{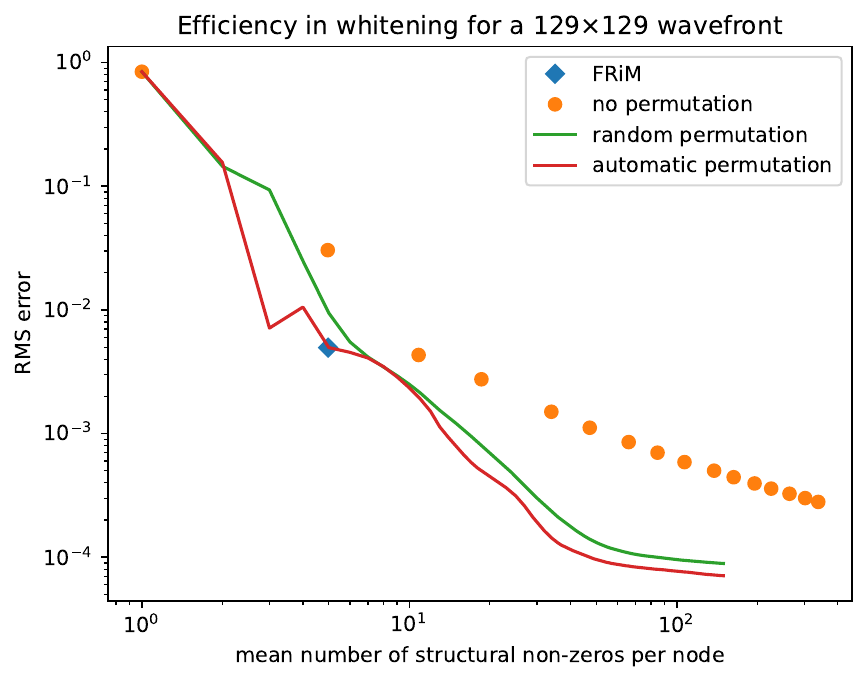}\hfill%
  \caption{Efficiency in whitening for wavefronts of size $17×17$, $33×33$,
    $65×65$, and $129×129$.  Four methods are compared: FRiM (blue diamond), a
    sparse structure with no permutations and a neighborhood of growing radius
    (orange circles), a sparse structure built for a random permutation and for
    a growing number of connected nodes (green line), and a sparse structure
    with an automatic ordering of nodes so as to uniformly sample the volume of
    the nodes at every stage (red line).  Abscissae are the mean number of
    structural non-zeros per node.  Ordinates are the RMS error:
    $\sqrt{\textfrac{1}{n}\,\Norm{\M K^{-1}⋅\M C⋅\M K\mT - \M I}_{\Tag{F}}^2}$
    with $n$ the number of wavefront nodes.}
  \label{fig:whitening}
\end{figure}

\section{RESULTS}
\label{sec:results}

To assert the ability of the different ASAP models to approximate the
statistics of the turbulence, we build different operators $\M K = \M P\T⋅\M
R^{-1}⋅\M P$ for different given permutations $\M P$ and various levels of
sparsity and compute the following root mean squared error:
\begin{equation}
  \mathrm{RMSE}(\M K) =
  \sqrt{\textfrac{1}{n}\,\Norm{\M K^{-1}⋅\M C⋅\M K\mT - \M I}_{\Tag{F}}^2}
\end{equation}
for $\M C$ the covariance matrix of a Kolmogorov wavefront, with $n$ the number
of wavefront samples, and where $\Norm{\ldots}_{\Tag{F}}$ denotes the Frobenius
norm.  Indeed $\M K^{-1}⋅\M C⋅\M K\mT$ would be the covariance matrix of the
whitened wavefront and thus, the closer is this covariance matrix to the
identity $\M I$, the better the approximation.  Figure~\ref{fig:whitening}
plots the value of $\mathrm{RMSE}(\M K)$ as a function of the mean number
$\mean{m}$ of structural non-zeros per node.  With $\mean{m} = 1$ all models
amounts to making a simple diagonal approximation for which $\mathrm{RMSE}(\M
K) \simeq 0.84$ hence a rather poor approximation as expected by the fact that
turbulent wavefronts have significant long range correlations.  Whatever the
model, the quality of the approximation improves as the mean number $\mean{m}$
of structural non-zeros per node grows.

For $\mean{m} \simeq 5$, FRiM's fractal structure always yields the lowest
errors.  Using the lexicographic order of the nodes (no permutation) is the
worst choice.  Taking a random permutation improves on the un-permuted case for
all $\mean{m}$.  These two methods are worse than FRiM's method for $\mean{m}
\simeq 5$. Finally, the proposed automatic ordering of the nodes to
approximately keep the same density of ordered nodes at any stage of the
construction yields the best results whatever $\mean{m}$, for $\mean{m} \simeq
5$ the quality of the approximation is as good as with FRiM.  These results
show that capturing the multi-scale structures of the covariance is highly
beneficial for the quality of the approximation and that the proposed method to
generalize the FRiM method is very effective.

\bibliography{asap} 
\bibliographystyle{spiebib} 

\end{document}

%% file: emmt-macros.tex
\usepackage{amsmath,amsfonts,amssymb}
\usepackage{mathtools}

\usepackage{graphicx}
\graphicspath{{figs/}}

\DeclareMathAlphabet{\mymathbb}{U}{BOONDOX-ds}{m}{n}

\usepackage[bb=ams]{mathalfa}

\usepackage[ruled,vlined]{algorithm2e}
\DontPrintSemicolon
\SetKw{KwAnd}{and}
\SetKw{KwOr}{or}

\SetKwComment{Comment}{\ensuremath{\vartriangleleft} }{}

\SetCommentSty{CommentFormat}
\newcommand{\CommentLine}[1]{{\normalsize\bf\ensuremath{\blacksquare} #1}\;}

\usepackage{xspace}

\usepackage[squaren,Gray]{SIunits}
\addunit{\flops}{flops}

\DeclareSymbolFont{bbold}{U}{bbold}{m}{n}
\DeclareSymbolFontAlphabet{\mathbbold}{bbold}
\RequirePackage{textcomp}  

\DeclarePairedDelimiterX{\Paren}[1]{(}{)}{#1}
\DeclarePairedDelimiterX{\Brace}[1]{\{}{\}}{#1}
\DeclarePairedDelimiterX{\Brack}[1]{[}{]}{#1}
\DeclarePairedDelimiterX{\Abs}[1]{\rvert}{\lvert}{#1}
\DeclarePairedDelimiterX{\Norm}[1]{\lVert}{\rVert}{#1}
\DeclarePairedDelimiterX{\Avg}[1]{\langle}{\rangle}{#1}
\DeclarePairedDelimiterX{\Round}[1]{\lfloor}{\rceil}{#1}
\DeclarePairedDelimiterX{\Floor}[1]{\lfloor}{\rfloor}{#1}
\DeclarePairedDelimiterX{\Ceil}[1]{\lceil}{\rceil}{#1}
\DeclarePairedDelimiterX{\Inner}[1]{\langle}{\rangle}{#1}
\DeclarePairedDelimiterX{\IntRange}[1]{\llbracket}{\rrbracket}{#1}
\DeclarePairedDelimiterX{\Group}[1]{\lgroup}{\rgroup}{#1}
\DeclarePairedDelimiterXPP{\List}[3]{}{\{}{\}}{_{#2,\ldots,#3}}{#1}

\newcommand*{\delimsize}{}
\newcommand*{\SuchThat}{\,\delimsize\vert\,} 
\newcommand*{\Given}{\,\delimsize\vert\,} 

\DeclareMathOperator*{\argmin}{arg\,min}

\DeclarePairedDelimiterXPP{\Var}[1]{\mathrm{Var}}(){}{#1}
\DeclarePairedDelimiterXPP{\Cov}[1]{\mathrm{Cov}}(){}{#1}
\DeclarePairedDelimiterXPP{\Expect}[1]{\mathrm{E}}(){}{#1}
\DeclarePairedDelimiterXPP{\LogPr}[1]{\ell}(){}{#1}

\newcommand*{\Set}[1]{\mathbb{#1}}
\newcommand*{\Reals}{\Set{R}}
\newcommand*{\Integers}{\Set{Z}}

\newcommand*{\from}{\,{:}\:}

\newcommand*{\Tag}[1]{\mathrm{#1}}

\newcommand*{\bydef}{\stackrel{{\scriptscriptstyle \text{def}}}{=}}

\newcommand*{\textfrac}[2]{{\textstyle\frac{#1}{#2}}}

\newcommand*{\V}[1]{\boldsymbol{#1}}   
\newcommand*{\M}[1]{\mathbf{#1}}      

\newcommand*{\Adj}{\top}
\newcommand*{\T}{^{\Adj}}
\newcommand*{\mT}{^{-\Adj}}

\renewcommand*{\cdot}{\mathord{\,\mathchar"2201\,}}

  \def\clap#1{\hbox to 0pt{\hss#1\hss}}


%% file: main.bbl
\begin{thebibliography}{1}

\bibitem{Thiebaut_Tallon-2010-FRiM}
Thiébaut, E. and Tallon, M., ``Fast minimum variance wavefront reconstruction
  for extremely large telescopes,'' {\em J.\ Opt.\ Soc.\ Am.\ A}~{\bf 27}(5),
  1046--1059 (2010).

\bibitem{Bechet_et_al-2007-Vancouver}
Béchet, C., Tallon, M., and Thiébaut, E., ``Closed-loop {AO} performance with
  fr{IM},'' in [{\em Adaptive Optics: Analysis and
  Methods}{\nolinebreak\hspace{0.1em}]},  {\em Conference of the Optical
  Society of America},  JTuA4 (2007).

\bibitem{Roddier-1990-wavefront_simulation}
Roddier, N., ``Atmospheric wavefront simulation using {Zernike} polynomials,''
  {\em Optical Engineering}~{\bf 29},  1174--1180 (Oct. 1990).

\bibitem{Lane_et_al-1992-Kolmogorov}
Lane, R.~G., Glindemann, A., and Dainty, J.~C., ``Simulation of a {Kolmogorov}
  phase screen,'' {\em Wave in random media}~{\bf 2},  209--224 (1992).

\bibitem{Bjorck-1996-least_squares}
{Å}ke Björck,  [{\em Numerical Methods for Least Squares
  Problems}{\nolinebreak\hspace{0.1em}]}, SIAM (1996).

\bibitem{Higham-2009-Cholesky_factorization}
Higham, N.~J., ``Cholesky factorization,'' {\em Wiley Interdisciplinary
  Reviews: Computational Statistics}~{\bf 1}(2),  251--254 (2009).

\end{thebibliography}
